\documentclass[twocolumn,tighten]{aastex63}

\hypersetup{linkcolor=magenta,citecolor=blue,filecolor=cyan,urlcolor=magenta}

\received{September 24, 2020}
\revised{October 23, 2020}
\accepted{October 26, 2020}

\submitjournal{ApJ}

\shorttitle{Coronal Inflows and In/Out Pairs}
\shortauthors{Lynch}

\begin{document}

\title{A Model for Coronal Inflows and In/Out Pairs}

\correspondingauthor{Benjamin~J.~Lynch}
\email{blynch@berkeley.edu}

\author[0000-0001-6886-855X]{Benjamin~J.~Lynch}
\affil{Space Sciences Laboratory, University of California--Berkeley,
        Berkeley, CA 94720, USA}











\begin{abstract}

This report presents a three-dimensional (3D) numerical magnetohydrodynamics (MHD) model of the white-light coronagraph observational phenomena known as coronal inflows and in/out pairs. Coronal inflows in the LASCO/C2 field of view (approximately $2-6\,R_\odot$) were thought to arise from the dynamic and intermittent release of solar wind plasma associated with the helmet streamer belt as the counterpart to outward-propagating streamer blobs, formed by magnetic reconnection. This interpretation was essentially confirmed with the subsequent identification of in/out pairs and the multispacecraft observations of their 3D structure. The MHD simulation results show relatively narrow lanes of density depletion form high in the corona and propagate inwards with sinuous motion which has been characterized as `tadpole-like' in coronagraph imagery. The height--time evolution and velocity profiles of the simulation inflows and in/out pairs are compared to their corresponding observations and a detailed analysis of the underlying magnetic field structure associated with the synthetic white-light and mass density evolution is presented. Understanding the physical origin of this structured component of the slow solar wind's intrinsic variability could make a significant contribution to solar wind modeling and the interpretation of remote and in-situ observations from \textsl{Parker Solar Probe} and \textsl{Solar Orbiter}. 

\end{abstract}


\section{Introduction} 
\label{sec:intro}

The launch of the \textsl{Solar and Heliospheric Observatory} \citep[SOHO;][]{Domingo1995} and the subsequent \textsl{Large Angle and Spectrometric Coronagraph} \citep[LASCO;][]{Brueckner1995} C2 and C3 white-light data ushered in an entirely new era of detailed observations of the structure and dynamics of the coronal streamers and the helmet streamer belt. \citet{Sheeley1997} analyzed the continual, intermittent outflow of density enhancements known as streamer blobs and the improved spatial and temporal resolution enabled the characterization of inflows in the wake of coronal mass ejections (CMEs) and streamer disconnection events \citep[e.g.,][]{WangYM1999a,WangYM1999b,WangYM2000}. Hundreds of coronal inflow events were observed and cataloged by \citet{Sheeley2002} and a new class of inflows in which an outward and inward component are clearly identified became known as in/out pairs \citep{Sheeley2007}. Coherent, small-scale flux rope structures were also being found in the in-situ observations from IMP~8 and \textsl{Wind} \citep{Moldwin2000} and were shown to have magnetic fields that could be fairly well-described by the same linear force-free cylinder model often used for larger interplanetary CME (ICME) flux rope structures \citep{Cartwright2008,Yu2014}. \citet{Crooker2004} analyzed in-situ observations of high-beta regions (heliospheric plasma sheets) in the \textsl{Wind} data that strengthened their interpretation of intertwined flux tubes, likely caused by interchange reconnection at the cusp of the helmet streamer belt \citep{Crooker1996hps,Crooker1996hfds}. 

The multispacecraft remote-sensing and in-situ observations provided by the \textsl{Solar Terrestrial Relations Observatory} \citep[STEREO;][]{Kaiser2008} spacecraft meant that \citet{Sheeley2009} were able to analyze the 3D structure of streamer blobs and under favorable spacecraft positions, the small flux ropes observed in the heliosphere were able to be directly traced back to their coronal source regions \citep[e.g.,][]{Kilpua2009b, Rouillard2010a, Rouillard2010b, Rouillard2011}. Advanced image processing enabled \citet{DeForest2012} to observe signatures of flux disconnection in the heliospheric imager field of view and \citet{HowardT2012}, \citet{SanchezDiaz2017b}, and others to better track and compile statistics of the kinematic evolution of both small-scale and large-scale transient outflows.

Recently, \citet{Hess2017inflows} have examined inflows in the inner white-light corona representing the closing down of magnetic flux beneath CME eruptions, \citet{WangYM2018blobs} have shown that gradual streamer expansion is often a precursor to streamer blob pinch-off at the cusp and whether an inflow is observed depends on the radial distance the reconnection occurs, which is in turn a function of field strength/flux content under the streamer belt, and \citet{SanchezDiaz2017a} have used STEREO data to argue that coronal inflows and streamer blob outflows are always associated, and in a follow up study, investigated the correspondence between small flux ropes, high density regions, and heliospheric current sheet (HCS) crossings in \textsl{Wind}, STEREO, and \textsl{Helios} data \citep{SanchezDiaz2019}. 

Data from the first several perihelia of \textsl{Parker Solar Probe} \citep[PSP;][]{Fox2016} already represent a treasure trove of new remote-sensing and in-situ observations: \citet{Wood2020} analyzed a streamer blob/disconnection event in the \textsl{Wide-field Imager for Parker Solar Pobe} \citep[WISPR;][]{Vourlidas2016} imaging data; \citet{Rouillard2020} have tracked helmet streamer outflow and its fluctuations in STEREO coronagraph and heliospheric imager data all the way to their in-situ measurement by PSP; and \citet{ZhaoLL2020} have shown the first PSP flyby contained multiple flux rope structures ranging in duration from 8--300~minutes. In addition, \citet{MurphyA2020} have recently analyzed small flux ropes in the solar wind seen with the MESSENGER spacecraft over a range of radial distances in preparation for better understanding current and future PSP and \textsl{Solar Orbiter} \citep[SolO;][]{Mueller2020} observations.

The underlying physics of streamer blob formation, inflows, and in/out pairs is more-or-less agreed upon, i.e.\ each of these signatures reflect coronal plasma dynamics resulting from magnetic reconnection associated with the open--closed field boundaries of coronal streamers, their evolution, and their intrinsic variability. However, the details of these reconnection processes, and our understanding of their role in creating the structured variability of the slow solar wind, remain an active area of research \citep[see reviews by][]{Abbo2016, Viall2020}.

%
%

%
{
Early attempts to model at least the outward moving part of these streamer disconnection/slow CME events were reasonably successful in 2.5D magnetohydrodynamic (MHD) simulations \citep[e.g.,][]{Linker1992} and confirmed that the dynamics associated with magnetic reconnection were, at least qualitatively, in agreement with the observations.  \citet{Suess1996} examined the role of coronal heating and heat conduction on the structure of the streamer belt, showing that the pointed, cusp-like feature was a result of the continual shedding of flux. \citet{Einaudi1999, Einaudi2001} and \citet{Rappazzo2005} modeled magnetic island formation in the wake--neutral sheet configuration of the streamer cusp--HCS system and \citet{Endeve2003, Endeve2004} characterized the lack of a stable equilibrium for the dipole streamer where the heating periodically accumulates enough gas pressure at the streamer cusp to overcome the magnetic tension forces causing the outermost layers to expand/open into the plasma sheet and solar wind. This expansion/opening facilitates magnetic reconnection in the equatorial current sheet that acts to closes magnetic flux back down, allowing the cycle to repeat.  \citet{ChenY2009} showed with a sufficient density gradient and velocity shear across the streamer boundary, a Kelvin--Helmholtz instability can develop and act as the gas pressure/mass density perturbation that drives magnetic reconnection at the cusp. Recently, \citet{Allred2015} have examined the force balance within 2.5D streamer blob plasmoids and shown their ejection periodicity can be controlled with the coronal heating factor.
}

%
\citet{Higginson2018} presented the first 3D MHD simulation of streamer blob formation at the cusp of the streamer belt and within the HCS in the extended corona for an idealized, solar minimum-like global field configuration. This paper continues the work of \citet{Higginson2018} through a detailed examination of the near-Sun consequences of magnetic reconnection occurring at the open--closed field interfaces of the coronal streamer belt in a more complex, solar maximum-like global field. The simulation results herein reproduce favorably many of the observed characteristics of coronal inflows, in/out pairs, and streamer blob flux rope formation, consistent with the \citet{Sheeley2002, Sheeley2007} and \citet{SanchezDiaz2017a} interpretations that all these phenomena result from the same magnetic reconnection processes occurring in the corona.

The paper is organized as follows. Section~\ref{sec:arms} presents a brief overview of the MHD model and the implementation of the initial magnetic field and solar wind boundary conditions. Section~\ref{sec:inflow} presents the simulation results including: (\ref{sec:inflow:wl}) the different types of coronal inflow morphologies in synthetic white-light emission; (\ref{sec:inflow:ht}) the analysis of the height--time and velocity profiles of the inward and outward moving transients; and an examination of the global (\ref{sec:inflow:mag}) and local (\ref{sec:inflow:mag2}) coronal magnetic field structure associated with these transient flows and their evolution. 
{
Section~\ref{sec:disc} discusses the applicability and extension of our simulation results to: (\ref{sec:disc:sads}) the low coronal supra-arcade downflows; (\ref{sec:disc:ps}) pseudostreamer outflows and reconnection dynamics; (\ref{sec:disc:ylm}) the relationship between inflows and the underlying magnetic field distribution; and (\ref{sec:disc:turb}) the structure of turbulence in the heliospheric plasma sheet. The summary and conclusions are presented in Section~\ref{sec:summary}.
}\vskip 0.10in

\begin{figure*}[!htb]
	\centering
	\includegraphics[width=0.95\textwidth]{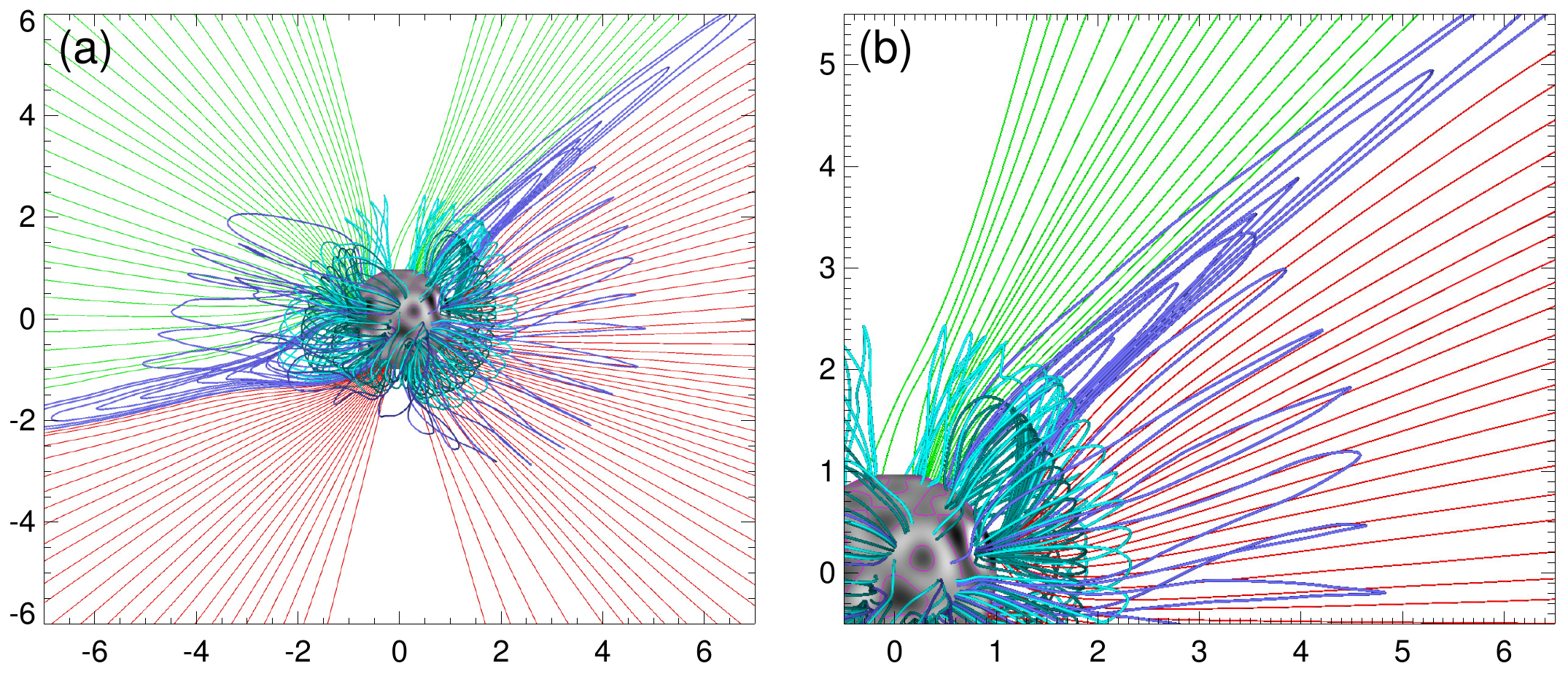}
	\caption{Magnetic field lines at $t=160$~hr after relaxing to a steady-state isothermal wind outflow. Panel (a): Approximately the LASCO C2 field of view with central meridian $\phi=60^{\circ}$ in Carrington longitude in CR2165. Positive (negative) polarity open fields are shown in green (red). The helmet streamer belt field lines are shown in the shades of blue. Panel (b): A closer-up view of the northwest quadrant to highlight the streamer belt structure near limb. The axis units are solar radii.
	}
	\label{fig:swfls}
\end{figure*}

\section{Quasi-steady State Solar Wind} 
\label{sec:arms}

The numerical simulation was performed with the Adaptively Refined MHD Solver \citep[ARMS;][]{DeVore2008} code. ARMS calculates solutions to the 3D nonlinear, time-dependent ideal MHD equations using a finite volume, multi-dimensional flux-corrected transport algorithm \citep{DeVore1991}. ARMS uses the PARAMESH framework \citep{MacNeice2000} for dynamic, solution-adaptive grid refinement and efficient multi-processor parallelization.

The spherical computational domain uses logarithmic grid spacing in $r$ and uniform grid spacing in $\theta, \phi$. The domain extends from $r \in [1\,R_\odot, 30\,R_\odot]$, $\theta \in [11.25^{\circ}, 168.75^{\circ}]$ ($\pm78.75^{\circ}$ in latitude), and $\phi \in [0^{\circ}, 360^{\circ}]$ (longitude). The initial grid consists of $7 \times 7 \times 15$ blocks with $8^3$ grid cells per block. There are 3 additional levels of static grid refinement and the level 3 refinement extends from $r \in [1\,R_\odot , 6.984\,R_\odot ]$ for all $\theta, \phi$. The level-4 grid refinement is centered on a southern hemisphere polarity inversion line for a separate study on the activation and eruption of a high-latitude filament.

The initial magnetic field configuration is constructed with a relatively low-degree ($\ell_{\rm max} = 14$) potential field source surface \citep[PFSS;][]{WangYM1992} extrapolation from the NSO/GONG \citep{Harvey1996} zero-point corrected, daily-updated $B_r$ synoptic map for Carrington Rotation 2165, taken on 2015 July~10 at 00:14UT.
%
%
A basic, quasi-steady state outflow is obtained via an isothermal \citet{parker1958} solar wind corresponding to a uniform temperature of $T_0 = 1.4 \times 10^{6}$~K \citep[e.g.,][]{Masson2013,Lynch2016b,Higginson2017a}. The base density is $n_0=3.62 \times 10^{8}$~cm$^{-3}$ at $1\,R_\odot$ and the initial radial velocity at the outer boundary is $v_{sw}(30\,R_\odot) \simeq 410$~km~s$^{-1}$. The magnetic field and outflow conditions adjust and eventually equilibriate creating the quasi-steady state open and closed flux distributions with a slow solar wind for $t > 100$~hr. The definition of `quasi-steady state' used here includes the small-scale, time-dependent dynamics of streamer evolution but maintains the stable, large-scale distribution of magnetic flux and resulting 3D solar wind structure in an average sense, along with essentially constant global energy measures (e.g. magnetic, kinetic, gravitaitonal, internal) in time.

\begin{figure*}
	\centering
	\includegraphics[width=1.0\textwidth]{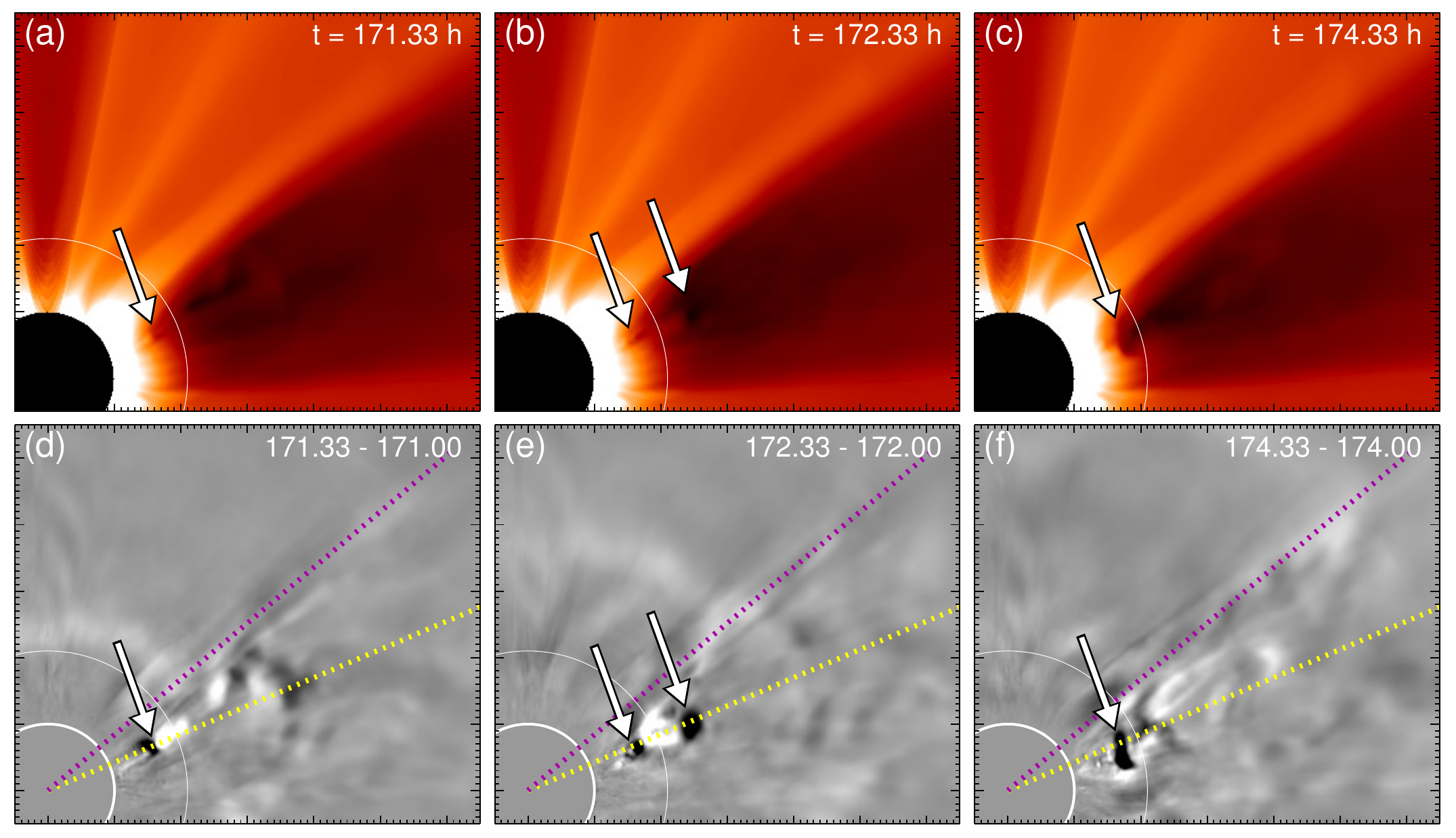}
	\caption{Representative examples of `sinking column' inflows along the PA 293$^\circ$ radial cut (dotted yellow line). Panels (a)--(c) show synthetic white-light coronagraph imagery from the MHD simulation. Panels (d)--(f) show the corresponding times in running-difference processing. Arrows indicate the leading edge of density depletion inflows. 
\\
	(An animation of this figure is available.)
	}
	\label{fig:wl}
\end{figure*}

Figure~\ref{fig:swfls}(a) shows the global coronal magnetic field during the quasi-steady state outflow at $t=160$~hr. Figure~\ref{fig:swfls}(b) plots a closer-up view of the northwest quandrant of panel (a) where we observe the simulation's coronal inflows. The axis ranges are normalized to solar radii. The same field lines are plotted in each panel: the open field lines are shown in green (red) for positive (negative) polarity; the closed field regions of the helmet streamer belt are illustrated with the blue field lines; and the set of dark cyan, cyan, and blue streamer arcade field lines are traced from points along the $B_r=0$ contour at heights $r=2.0R_{\odot}$, $2.5R_{\odot}$, and $5.0R_{\odot}$, respectively. Additional blue field lines are plotted above the streamer belt at each limb (i.e. in the plane of the sky) to indicate the global orientation in the extended corona. The east limb is at $\phi = -30^{\circ}$ and the west limb at $\phi = 150^{\circ}$ longitude. The spatial dimensions of Figure~\ref{fig:swfls}(b) correspond to the exact axis ranges for the panels in Figures~\ref{fig:wl} and \ref{fig:wl2}, and further analysis of the helmet streamer magnetic structure is presented in $\S\S$\ref{sec:inflow:mag}, \ref{sec:inflow:mag2}.

\section{Inflows and In/Out Pairs} 
\label{sec:inflow}

\subsection{Morphology in Synthetic White-light Images}
\label{sec:inflow:wl}

\citet{Sheeley2002} described the morphology of the most common type of coronal inflow in coronagraph images as a `sinking column' in which a weak localized density enhancement appears between 3—$5\,R_\odot$ and accelerates towards the sun (and then decelerates) while leaving a dark, collimated, and extended channel structure in its wake, corresponding to a 10--30\% intensity depletion. These inflows are sometimes referred to as `raining inflows.' Figure~\ref{fig:wl} plots the synthetic Thomson scattered white-light intensity from the 3D MHD data cube of number density assuming $n_p = n_e$ \citep[as in, e.g.,][]{Lynch2004,Lynch2016b,Vourlidas2013}. The top row, panels (a)--(c), show the ratio $I(t)/I_0$ at times 171.33~hr, 172.33~hr, and 174.33~hr, where $I_0$ is obtained from the $t=0$~hr spherically symmetric density profile. The bottom row, panels (d)--(f), show the running-difference processing of the synthetic white-light images above, defined as $\Delta I = \left( I(t) - I(t-\Delta t) \right)/I_0$ with $\Delta t = 20$~min. 

\begin{figure*}
	\centering
	\includegraphics[width=1.0\textwidth]{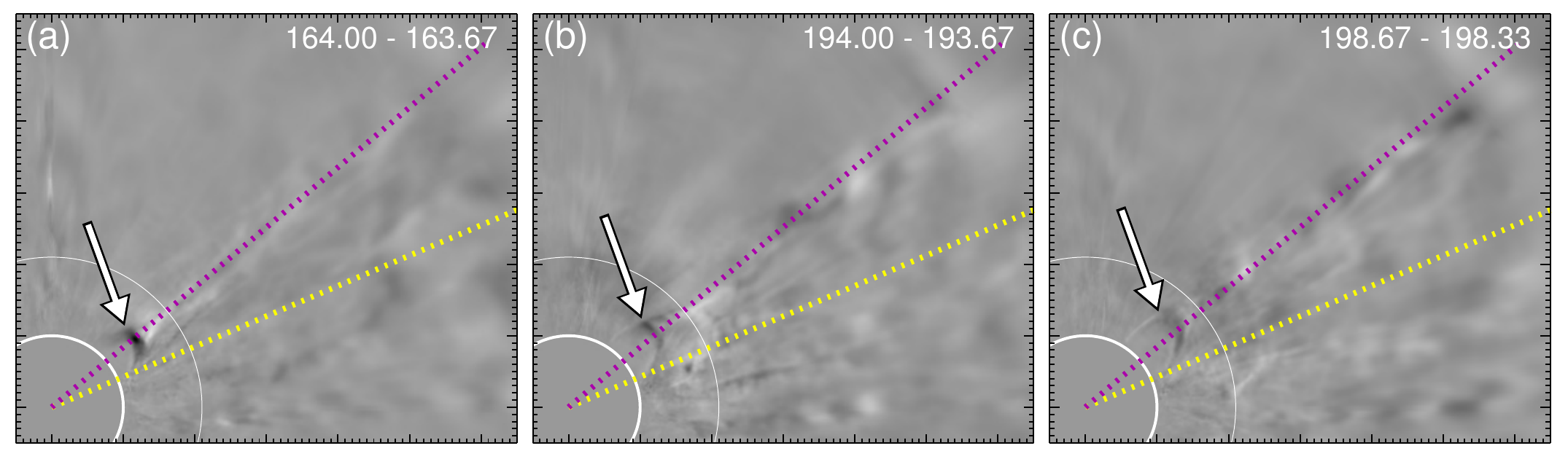}
	\caption{Representative examples of `shrinking loop' inflows along the PA 310$^\circ$ radial cut (dotted purple line) in the same format as Figure~\ref{fig:wl}(d)--(f). Arrows indicate the darker leading loop and brighter training cusp. 
	}
	\label{fig:wl2}
\end{figure*}

The arrows in Figure~\ref{fig:wl}(a)--(c) show the location of the leading edge of the dark `sinking columns' characteristic of coronal inflow observations. The same arrows are also in Figure~\ref{fig:wl}(d)--(f) as pointing to the clear bipolar intensity signals of the coronal inflows above. The morphology of these simulation features is essentially identical to that of observed inflows (e.g., cf. Figure~3 in \citealt{WangYM1999a}; Figure~9 in \citealt{WangYM2000}; Figures~1, 4, and 7 in \citealt{Sheeley2002}; Figures~7 and 8 in \citealt{Sheeley2004}). The animated version of Figure~\ref{fig:wl} highlights the dynamical evolution of the synthetic white-light and running-difference features.

The second type of inflow has a morphology described as a `shrinking loop' corresponding to the downward motion of a dark, arched loop structure, often accompanied by a trailing brighter cusp shape in the running-difference processing. In the standard background-subtracted white-light intensity images, these types of inflows are much harder to see, i.e. often there is just a slight downward motion/contraction of a semi-circular contour of streamer brightness. Figure~\ref{fig:wl2} shows three representative examples of shrinking loop inflows (indicated by the arrows) in the same format as the Figure~\ref{fig:wl} running-difference panels. The morphology of these simulation features is again, essentially identical to this type of observed inflow as well (e.g., cf. Figure 7 in \citealt{WangYM1998}; Figure 9 in \citealt{Sheeley2007}; Figures 1, 4, and 10 in \citealt{Hess2017inflows}; Figures 1 and 3 in \citealt{WangYM2018blobs}).

Figures \ref{fig:wl}(d)--(f) and \ref{fig:wl2} also have radial cuts indicated at position angles (PAs) 310$^\circ$ (purple dotted line) and 293$^\circ$ (yellow dotted line) corresponding to latitudes of +40$^\circ$ and +23$^\circ$, respectively. These radial cuts are used to construct the height--time evolution of the outflow and inflow running-difference intensity features in the next section. It is also worth noting that the sinking-column inflows are located at the PA 293$^\circ$ while the shrinking-loop inflows appear at PA 310$^\circ$. Section~\ref{sec:inflow:mag} will show these PAs correspond to viewing the helmet streamer belt edge-on (PA 310$^\circ$) and face-on (PA 293$^\circ$).

\begin{figure*}
	\centering
	\includegraphics[width=1.0\textwidth]{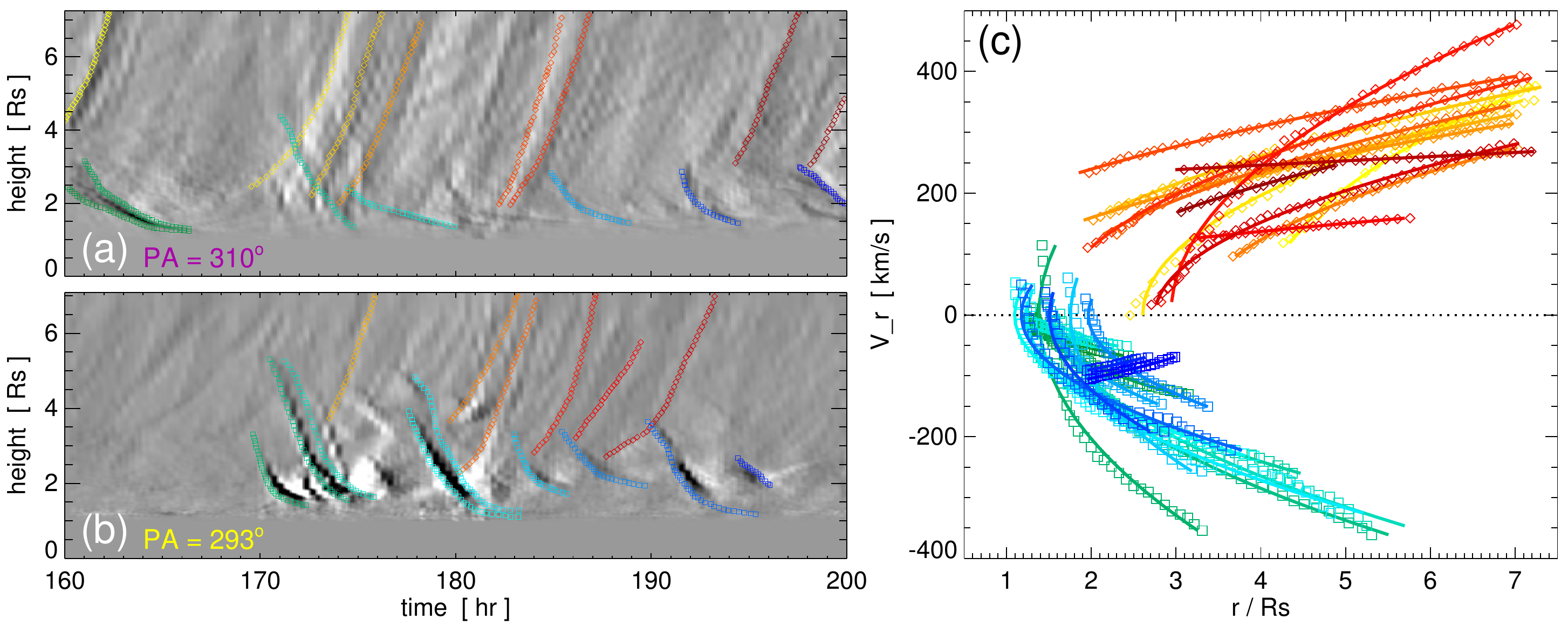}
	\caption{height--time plots for the two radial cuts in Figures~\ref{fig:wl}, \ref{fig:wl2} and their resulting velocity profiles. Panel (a): PA 310$^{\circ}$ samples an edge-on portion of the helmet streamer belt. Panel (b): PA 293$^{\circ}$ samples a face-on portion of the helmet streamer belt. Panel (c): Each of the $v_r(r)$ profiles derived from the quadratic fits to the height--time data in (a), (b). The inflow (outflow) tracks are shown in the green-cyan-blue (yellow-orange-red) color gradient. 
	}
	\label{fig:jmap}
\end{figure*}

\subsection{Height--time and Velocity Profiles}
\label{sec:inflow:ht}

To quantify the apparent motion of the inwards and outwards moving features in the synthetic white-light running difference movie, height--time plots (`J-maps') are constructed at the two position angles indicated in Figures~\ref{fig:wl} and \ref{fig:wl2}. Figures~\ref{fig:jmap}(a) and \ref{fig:jmap}(b) show the running difference height--time plots from PAs 310$^{\circ}$, 293$^{\circ}$, respectively. A set of height--time tracks are traced via the point-and-click method and shown as yellow-orange-red data points for the outflows and green-cyan-blue for the inflows. The height--time data are fit using the IDL {\tt curvefit} procedure to the standard quadratic profile \citep[as in][]{Sheeley1997} given by
$r(t) = r_0 + v_0 t + \onehalf a_0 t^2$
resulting in a velocity profile of $v^2(r) = 2 a_0 ( r - r_1 )$ where $r_1 = r_0 - v_0^2/(2a_0)$.

Figure~\ref{fig:jmap}(c) plots the $v_r(r)$ profiles from each of the quadratric fits to the height--time data in panels (a) and (b) in the same color scheme. For the most part, the set of inflow velocity profiles and the set of outflow velocity profiles are each relatively consistent---there is some variation between tracks with the occasional outlier, but overall, each set is essentially clustered together within a $\pm$100~km~s$^{-1}$ envelope. The inflow tracks originating at the greatest radial distances start with an initial negative velocity between $-200$ and $-350$~km~s$^{-1}$ for $r > 3\,R_\odot$ and rapidly decrease in velocity as they approach the Sun; e.g. for $r < 2\,R_\odot$, most of the inflow velocities are slower than $-100$~km~s$^{-1}$ and many of the analytic fits overshoot the $v_r = 0$ threshold by the last few height--time points. The outflow tracks show a broader distribution in their initial $v_r$ at lower radial distances ($r < 3\,R_\odot$) but they tend to narrow with distance until the upper boundary of the height--time plots; $v_r(7\,R_\odot) \sim 325\pm75$~km~s$^{-1}$.

The simulation inflow and outflow velocity magnitudes and their radial dependence are reasonably consistent with the observed profiles. Typical inflow velocities are observed to reach a maximum speed of approximately $-100$~km~s$^{-1}$ (e.g., cf. Figure~5 in \citealt{WangYM1999a}; Figure~2 in \citealt{WangYM1999b}). The observed streamer blob outflow velocity profiles are usually clustered around the ambient slow solar wind profile and at $r=7\,R_\odot$ these are in the $200\pm100$~km~s$^{-1}$ range (e.g., cf. Figure~6 in \citealt{Sheeley1997}; Figure~7 in \citealt{WangYM2000}; Figure~5 in \citealt{Song2009}). The simulation inflow velocity profiles tend to start a bit larger than the initial velocity magnitudes in the observations, but they rapidly decelerate to radial velocity magnitudes comparable to observed coronal inflows close to the sun (e.g. for $ r \lesssim 2\,R_\odot$). Likewise, the streamer blob/density enhancement outflows tend to start and remain a bit faster than the 400~km~s$^{-1}$ isothermal slow solar wind profile. Given that the inflow tracks and most of the outflow tracks originate in reconnection exhaust, it is not surprising the simulation and observed velocity profiles are not an exact match.

\subsection{Global Magnetic Field Structure and Dynamics}
\label{sec:inflow:mag}

Some of the first statistical results from coronal inflow observations was their occurance frequency followed the solar activity cycle \citep{WangYM1999a,Sheeley2002} and they almost always appeared in regions of the corona associated with sector boundaries \citep{Sheeley2002,Sheeley2007}, i.e. the transition from one open field polarity to the other across the helmet streamer belt and HCS. The orientation of the HCS changes drastically with the solar cycle, and solar maximum magnetic field configurations often have large latitudinal excursions of the helmet streamer belt. Since the coronal inflows are best observed in these highly distorted/vertical sections of the helmet streamer belt (i.e. when the HCS is parallel to the plane of the sky), then the relationship between observed inflow occurance to solar activity is straightforward.

\begin{figure}
	\centering	\includegraphics[width=0.46\textwidth]{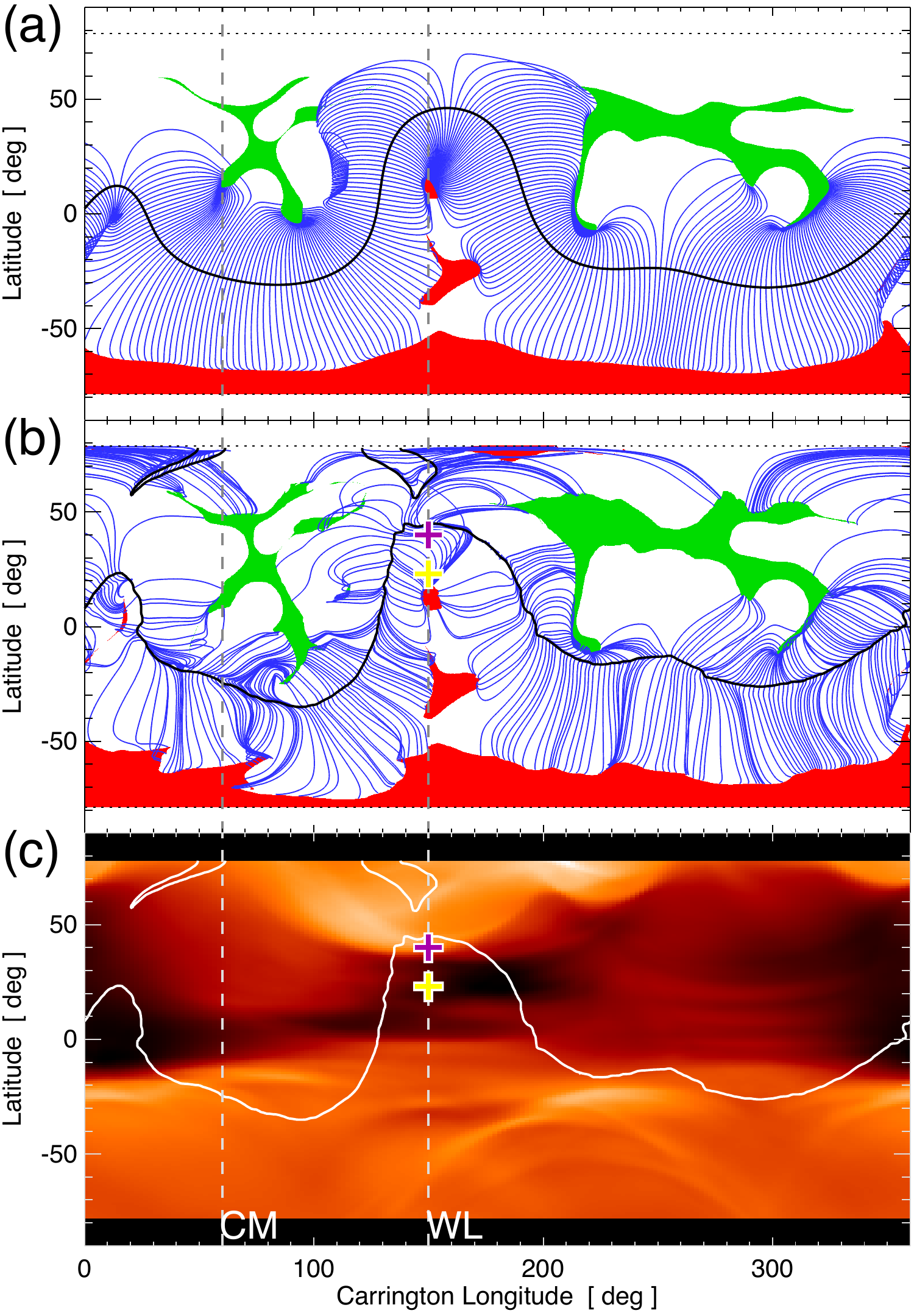}
	\caption{Panel (a): The initial $t=0$~hr PFSS magnetic field derived from NSO/GONG synoptic map for Carrington Rotation 2165. The positive (negative) open field regions are indicated in red (green), the HCS neutral line is shown as the black contour and the helmet streamer belt field lines are shown in blue. Panel (b): The $t=160$~hr global magnetic field associated with the quasi-steady state solar wind outflow in the same format as above. Panel (c): Synthetic line-of-sight integrated white-light emission at $5\,R_\odot$ as a synoptic map. The central meridian (CM) and west limb (WL) for Figure~\ref{fig:wl} are shown as the vertical dashed lines in each panel. The purple and yellow `+' signs indicate the WL position angles of the radial cuts used for the height--time plots in Figure~\ref{fig:jmap}.   
	}  
	\label{fig:mag1}
\end{figure}

To investigate the global coronal context for the various simulation inflows, Figure~\ref{fig:mag1} shows the large-scale magnetic field and streamer belt configuration in Carrington rotation coordinates. Figure~\ref{fig:mag1}(a) is PFSS reconstruction from the NSO/GONG radial field synoptic map in the style of the GONG data products: Positive open field regions are green, negative open field regions are red, the boundary of the helmet streamer belt is in blue, and the black $B_r=0$ neutral line at $r=2.5\,R_\odot$ indicates the base of the HCS. Figure~\ref{fig:mag1}(b) shows the MHD version of the panel (a) figure at $t=160$~hr. Overall, the large-scale coronal structure in the presence of the isothermal solar wind maintains an excellent qualitative agreement to the PFSS extrapolation. Figure~\ref{fig:mag1}(c) plots the synoptic map of white-light intensity in Carrington coordinates. The synoptic map was created by generating 90 synthetic coronagraph images in 2$^{\circ}$ increments from the $t=160$~hr data cube and then sampling each one along the $r = 5\,R_\odot$ circle and assigning each limb to their corresponding Carrington longitudes. The $B_r=0$ neutral line is also overplotted. The white-light Carrington map is a standard procedure employed in the analysis of coronagraph data \citep[e.g.,][]{WangYM1999b} and has been used recently by \citet{Rouillard2020} in linking streamer outflows to PSP in-situ observations.

The west limb (plane of the sky in Figure~\ref{fig:wl}) is indicated as `WL' at Carrington longitude $\phi=150^{\circ}$ and the central meridian is labeled `CM' at $\phi=60^{\circ}$. The latitude and longitude position of the radial cuts in Figures~\ref{fig:wl}, \ref{fig:wl2} are shown in Figure~\ref{fig:mag1}(b) and \ref{fig:mag1}(c) as the purple and yellow `+' symbols. The streamer belt/HCS is highly inclined and in a face-on orientation 10--15$^{\circ}$ in front of the plane-of-they-sky (at $\phi \sim 140^{\circ}$). The streamer belt then wraps around above the radial sampling points to become edge-on at high latitude (HCS in the $r$--$\phi$ plane at $\theta \sim 40^{\circ}$) and comes back down on the other side of the negative polarity coronal hole extension---this time 30--40$^{\circ}$ behind the plane-of-the-sky (longitude $\phi \sim 190^{\circ}$).

The large-scale, global context provided by Figure~\ref{fig:mag1} is especially important when trying to unravel the contributions of different coronal structures to the line-of-sight integration. For example, the upper radial sample (purple `+') occurs in the dense streamer stalk outflow and has approximately $\pm25^{\circ}$ of streamer outflow material along the line of sight at $r=5\,R_\odot$ centered on the plane of the sky. Conversely, the lower radial sample (yellow `+') is above the extension of the negative polarity coronal hole which cannot contribute much to the white-light intensity because of the low density on open field lines, despite the favorable scattering geometry. Rather, the significantly more dense helmet streamer arcade and HCS extension only 10--15$^{\circ}$ east from the plane of the sky is largely responsible for the observed inflow and outflow dynamics.

\begin{figure*}
	\centering
	\includegraphics[width=1.0\textwidth]{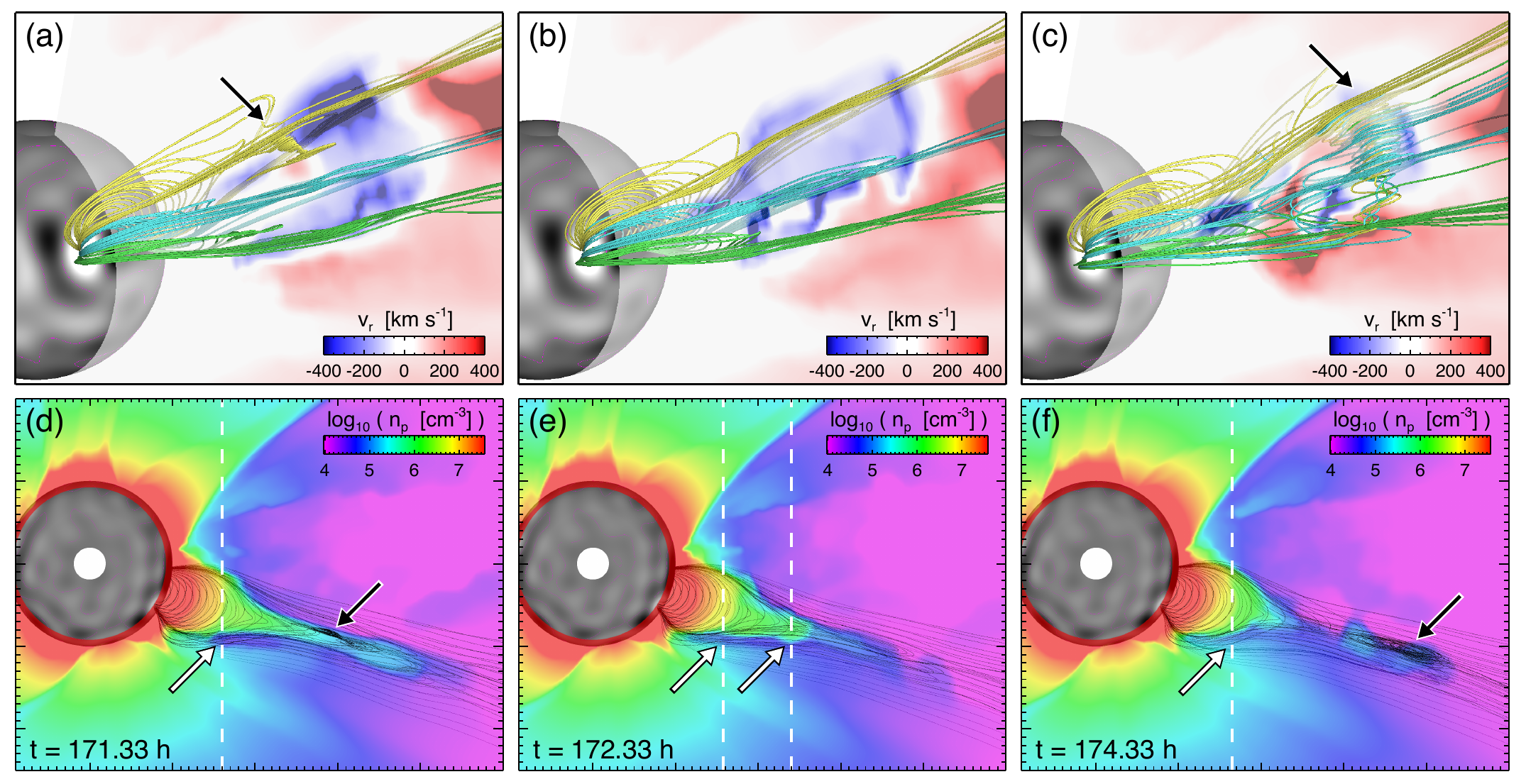}
	\caption{Visualization of the magnetic field structure and density distribution during the inflow events of Figure~\ref{fig:wl}. Panels (a)--(c): 3D perspective of representative magnetic field lines of the streamer belt and the base of the HCS. The semi-transparent meridional plane bisecting the streamer arcade at $\phi=135^{\circ}$ shows $v_r(r,\theta)$ in red/blue for the $\pm{\hat{r}}$ directions. Panels (d)--(f): The density distribution $n(r,\phi)$ in a semi-transparent latitudinal plane at $\theta=+23^{\circ}$ viewed from the solar north pole. The dashed white lines show the lines of sight at the positions of the `sinking column' inflows in Figure~\ref{fig:wl} and the white arrows indicate their intersection with the collimated density-depleted legs of newly reconnected loops retracting back down. The black arrows indicate magnetic island plasmoids formed by reconnection at the streamer cusp/HCS base. \\
	(An animation of this figure is available.) 
	}
	\label{fig:mag2}
\end{figure*}

\subsection{Local Magnetic Field Structure and Dynamics}
\label{sec:inflow:mag2}

Figure~\ref{fig:mag2} shows the magnetic field and plasma evolution at the vertical/warped portion of the helmet streamer belt near the west limb. Panels \ref{fig:mag2}(a)--(c) show a 3D perspective of a transparent meridional plane of radial velocity at Carrington longitude $\phi=135^\circ$ at the three times shown in Figure~\ref{fig:wl} ($t=171.33$, 172.33, and 174.33~hrs). A set of magnetic field lines are traced from starting points in the $\phi=135^{\circ}$ plane at latitudes $+18^\circ$ (green), $+23^\circ$ (blue), and $+30^\circ$ (yellow) over a range of radial distances to make up three representative arcades in the vicinity of the PA 293$^\circ$ (23$^\circ$ latitude) radial cut used to generate up the height--time plot in Figure~\ref{fig:jmap}(b). The blue--white--red gradient in radial velocity above the streamer arcade is indicative of the outflows caused by localized reconnection at the extended cusp of the streamer and base of the HCS. Here the negative radial velocities (blue) show the reconnection exhaust flowing back towards the sun. The animation of Figure~\ref{fig:mag2} makes clear the transient nature of the reconnection in this region---it moves around, the outflow velocities increase and decrease in intensity, and in general, the reconnection sets in after sufficient accumulation of material and stretching of the arcade field lines into the HCS region. The reconnection at the streamer cusp/HCS base forms 3D plasmoid flux ropes that are either ejected into the dense neutral sheet surrounding the HCS or are ejected back into the streamer arcade flux system. The black arrows in panels \ref{fig:mag2}(a) and \ref{fig:mag2}(c) point to two of these reconnection flux ropes; the first one is clearly in the downflow and rejoins/reconnects back into the streamer arcade while the second one is in the upflow and becomes part of the dense plasma sheet outflow.

Figure \ref{fig:mag2}(d)--(f) shows the number density on a transparent plane at latitude $\theta = 23{^\circ}$ (again, corresponding to the yellow radial cut in Figures~\ref{fig:wl}, \ref{fig:wl2}). The vertical dashed lines show the position along the $x$-axis corresponding to the location of the arrows in Figure~\ref{fig:wl}. The white arrows in panels \ref{fig:mag2}(d)--(f) indicate the narrow, newly-reconnected, evacuated outer layer of the helmet streamer. In the Figure~\ref{fig:mag2} animation, these low-density columns are seen moving towards the sun, tracing both edges of the helmet streamer along the magnetic field structure. These tenuous downflows at the flanks of the streamer arcade when viewed edge-on originate in the reconnection exhaust downflows incident at the streamer cusp shown above that then flow down the legs of the arcade loops. The same magnetic island plasmoid flux ropes from \ref{fig:mag2}(a) and \ref{fig:mag2}(c) are highlighted with black arrows in \ref{fig:mag2}(d) and \ref{fig:mag2}(f). The magnetic island plasmoid flux ropes are always associated with a localized density enhancement in the current sheet \citep[see also][]{Lynch2016a}. The sunward-moving plasmoids are inflowing density enhancements that reconnect with and rejoin the coronal arcade while the outflowing plasmoids correspond to the classic streamer blob density enhancements with a 3D magnetic flux rope structure.

The animation of Figure~\ref{fig:mag2} shows both of the physical processes proposed by \citet{WangYM2000} for inflows: contraction of loops after reconnection---a non-energetic, large-scale version of the geometry of flare reconnection and post-eruption arcade formation from a tiny pinch-off flux rope high in the corona, and the retraction of stretched/distended loops via magnetic tension and gravity.
The expansion--reconnection--contraction--expansion cycle results from the interplay between the gas pressure and magnetic field tension in the streamer cusp region \citep[e.g.,][]{Endeve2004,ChenY2009,Allred2015}. One of the reasons the cusp of the helmet streamer belt is so ``unsteady'' is the difference in the response to small perturbations between an X-point null and a Y-point null. Reconnection at an X-point null driven by a short-duration, external transient can cause oscillatory reconnection \citep[e.g.,][]{McLaughlin2009,Murray2009} that will damp out because the different flux systems can transfer flux back and forth via reconnection until the magnetic stresses have dissipated, whereas the Y-point null (line) at the streamer belt cusp and base of the HCS is less stable. The plasma sheet region has a significantly lower magnetic pressure than the open field on either side (and $B \sim 0$ in the current sheet itself). Thus, any pressure fluctuations from the streamer large enough to overcome the magnetic tension of the outer flux surfaces will just move the Y-point, effectively causing the streamer flux system to expand. Once reconnection has set in at the over-expanded cusp, the streamer flux system is reconfigured so quickly that it overshoots the force-balance equilibrium and the cycle can begin again. The animation of Figure~\ref{fig:mag2} illustrates exactly this process.

\section{Discussion}
\label{sec:disc} 

\subsection{Extension to Supra-Arcade Downflows}
\label{sec:disc:sads}

Once coronal inflows had been identified and characterized in the coronagraph observations, there was an considerable effort to figure out if the same physical processes and evolution of the magnetic field were responsible for the phenomena known as supra-arcade downflows \citep[SADs;][]{McKenzie2000, McKenzie2009, Savage2010, Savage2011, Savage2012}. SADs are observed much lower in the corona in EUV and X-ray measurements of the hot plasma surrounding the reconnecting current sheet above post-eruption arcades in the aftermath of CME eruptions. SADs are almost identical to the `sinking column' coronal inflows in their shape, radial and transverse motions, and in the trailing dark lanes in their wake \citep[e.g.,][]{Sheeley2004}. \citet{Savage2010} showed that the deprojected SADs in the current sheet from $r \le 1.42\,R_\odot$ had initial velocities between $0$ and $-200$~km~s$^{-1}$. We note that, with one exception, all of the Figure~\ref{fig:jmap} inflow tracks have radial velocities below $-200$~km~s$^{-1}$ for $r < 2\,R_\odot$. SADs are most often seen when the flare arcade is oriented such that the current sheet above the arcade loops appears face-on. This is an equivalent viewing orientation as when the helmet streamer belt/HCS is significantly warped with a large latitudinal extent, i.e. at the sector boundaries (such as the streamer orientation in Figure~\ref{fig:mag2}).

\citet{Cassak2013} performed numerical simulations to examine the relationship between SADs and flare reconnection outflows. Their model was based on (1) a realistic density stratification so the less-dense reconnection jet outflow creates a depletion, (2) the reconnection being steady enough and of sufficient duration to keep plasma from filling this depletion in, and (3) localization of the reconnection site with respect to the length of the current sheet for the jet outflow to remain collimated. This was in contrast to the intermittent, patchy reconnection models proposed earlier \citep[e.g.,][]{Linton2006, Longcope2010} which were also able to reproduce some qualitative agreement with properties of observed SADs, such as the height--time profile and trailing density voids \citep{Linton2009,Guidoni2011}.

The results presented herein are, in some sense, a mix of the \citet{Cassak2013} and \citet{Linton2009} scenarios. The global corona modeled here has both gravitational stratification as well as a significant density variation between the open and closed flux systems, meaning the open field lines being swept into the HCS dissipation region to form newly-closed loops that retract back down, have a much lower (open-field) mass density than the rest of the underlying (closed-field) streamer system. However, the reconnection at the streamer cusp/HCS base is also fairly ``patchy.'' The formation and ejection of magnetic island plasmoids that transfer the mass and magnetic flux into the streamer arcade take the form of time-dependent, bursty reconnection jet outflows rather than a continuous smooth and stable outflow profile. Consequently, the trailing dark tails seen in the face-on PA~293$^{\circ}$ location are relatively short-lived in these simulation results.

A similar effect was seen by \citet{Edmondson2017} during their investigation of the density fluctuations in plasmoid-unstable current sheets with varying guide field strengths. In those simulations, the reconnection outflow into the denser, closed-flux region produced density voids with the same `tadpole-like' morphology of a `sinking column' with a dark, sinuous wake (see the mass density panels in Figures 6--8 of \citealt{Edmondson2017}). The most visible (darkest) and collimated reconnection jet outflows were in the zero guide field case, they were a little broader and still visible at 10\% guide field, and they were significantly wider and less visible in the 50\% guide field case (i.e. less contrast with respect to background density). When viewing the current sheet face-on, the spatial width of these intermittent low-density reconnection jet outflows appears to increase in proportion to the strength of the guide field component. The \citet{Edmondson2017} results provide an intuitive explanation for why SADs are more visible in flare arcade plasma sheets during the gradual or decay phase of long-duration events \citep{McKenzie2000}. During the impulsive phase of a flare there is a strong guide field component in the reconnection region from the highly sheared and twisted fields of the erupting structure. Therefore, the density-depleted outflows are more spread out (less collimated) parallel to the current sheet. By the late gradual/decay phase of the flare, there is not much guide field left to reconnect in the wake of a CME \citep[e.g. Figure 11 in][]{Lynch2016a}. At this stage the post-eruption arcade flare loops are reforming with significantly less shear \citep{Aulanier2012} and the density-depleted outflows are more aligned (collimated) perpendicular to the post-eruption arcade current sheet.

While it is relatively well accepted that inflows in white-light coronagraph data, especially in the wake of CMEs, are just larger versions of SADs \citep[e.g.,][]{Sheeley2004}, the simulation results presented herein also highlight the importance of the viewing perspective in determining the apparent structure and morphology of the inflows. For the common `raining' or `sinking column' inflows at sector boundaries, the line-of-sight integration is likely to intersect both legs of the density-depleted, retracting loop if the streamer belt (or flare arcade) is oriented such that the HCS (flare current sheet) is parallel to plane of sky and has a negligible guide field component---exactly the case as in Figure~\ref{fig:mag2}(d)--(f). If there is still a large-scale shear component or the arcade is oriented at a significant angle with respect to the plane of the sky, then the smaller contribution to the line-of-sight integral will result in less visible inflow signatures. The `shrinking loop' inflows seen in the edge-on streamer at PA 310$^{\circ}$ result from the same retraction of a density-depleted loop process as depicted in the viewpoint of Figure~\ref{fig:mag2}(d)--(f). Here the perspective from the solar north pole views the vertical, warped portion of the streamer belt (at $\phi=135^{\circ}$) as edge-on and perpendicular to the line of sight for Figures~\ref{fig:wl}, \ref{fig:wl2}.

\subsection{{Extension to Pseudostreamer Outflows and Reconnection Dynamics}}
\label{sec:disc:ps} 

{
The magnetic topology of coronal pseudostreamers has been discussed in detail \citep[e.g.,][]{WangYM2007a, WangYM2012, Titov2012, Panasenco2013, Rachmeler2014, Scott2018} along with observations and modeling of their slow solar wind outflow \citep{Crooker2012a, Riley2012, WangYM2012, Owens2014}. Pseudostreamers differ from helmet streamers in that they are surrounded by a single open field polarity rather than separating positive and negative polarity open fields which creates a HCS and sector boundary. The network of pseudostreamer connectivity in the heliosphere and its relation to the main helmet streamer belt has been named the Separatrix Web \citep[S-Web;][]{Antiochos2011,Antiochos2013} and is a favorable location for interchange reconnection \citep[e.g.,][]{Higginson2017a}. 
}

{
\citet{Masson2014} examined gradual reconnection in a slowly stressed, 3D separatrix fan-spine configuration showing a smooth, continuous outflow of material along the external spine line. \citet{Lynch2013} showed that pseudostreamer interchange reconnection, in the form of pre-eruption breakout reconnection, could result in bursty but quasi-steady signatures in density along the external spine and coronal dimming signatures near the stressed null point and current sheet \citep[e.g.,][]{KumarP2020}. In that simulation the sunward portion of the interchange reconnection outflow became downflows in the adjacent flux system loops, similar to the dynamics in Figure~\ref{fig:mag2}. Another type of downflow is `coronal rain' which is also observed at null points in multipolar systems \citep[e.g.,][]{Mason2019}. This type of downflow is primarily a result of plasma thermodynamics (thermal nonequilibrium condensation of cooler, dense material observed in EUV, e.g. SDO/AIA 304~{\AA}), although interchange reconnection may also play some role in its subsequent transport.
}

{
Streamer blobs originating from pseudostreamers are far less common and/or visible in white-light coronagraph observations \citep[e.g.,][]{WangYM2007a,WangYM2012}. However, there are some in-situ observations of bidirectional electron signatures that are suggestive of closed-field structures with coronal connections at both foot points and a subset of these interplanetary small-scale flux ropes appear to originate from coronal pseudostreamers, i.e. far from the HCS and its plasma sheet \citep[e.g.,][]{Feng2015b}. Future numerical simulations will be required to characterize pseudostreamer wind variability and its comparison with helmet streamer slow wind.}

\subsection{Relating Inflow Occurrence to the Underlying Magnetic Field Distribution}
\label{sec:disc:ylm} 

{
As discussed above, the interplay between gas dynamics and the magnetic field at the Y-type null line at the cusp of helmet streamers is ultimately responsible for the ``unsteady'' character of the quasi-steady state outflows. Any change in the relative balance of forces will cause a similar cycle of expansion, reconnection and over-correction followed by another period of expansion in response to the over-correction. For example, an increase in the magnetic pressure of the closed flux system through the addition of new flux (through photospheric emergence) or the addition of magnetic stresses through coherent surface motions (large-scale shearing, or other global flow patterns such as differential rotation) or incoherent surface motions (convective turbulence/granulation and/or the accumulation of large-scale twist via helicity condensation) could result in a similar disruption.
}

{
As the streamer flux system responds to evolving surface fields, the streamer swells in width and expands radially. \citet{WangYM2018blobs} have examined this process in detail and shown the occurrence of coronal inflows are well-correlated with the total polarized brightness ($pB$ radiance) in the LASCO C2 field of view on the timescale of Carrington rotations. On a global scale, the increase in the number of sector boundaries---the vertical, highly warped portions of the helmet streamer belt---occurring in the ecliptic plane can also be represented by the power (coefficient magnitude) in the nonaxisymmetric components of the PFSS spherical harmonic expansion ($Y_{l,|m|}$). In particular, the equatorial dipole ($Y_{1,1}$) and equatorial quadrupole ($Y_{2,2}$) components are known to vary with the solar activity cycle \citep{WangYM1997} and thus, also show good correlation with the observed inflow rates \citep[e.g.,][]{Sheeley2014}. 
}

{
\citet{WangYM2018blobs} showed that the highest inflow rate and C2 $pB$ radiance levels in cycle 24 were recorded during 2014 October--December, above NOAA 12192 and its remnants. During this period, the helmet streamer overlying active region 12192 appears to have expanded well beyond its ``normal'' position. This sunspot grouping was extremely active, generating many strong flares (6 X- and $\sim$18 M-class flares) that, surprisingly, only resulted in a single CME eruption during the period October 18--28 \citep[e.g.,][]{SunX2015}. This (confined) flaring activity would certainly produce an abundance of significant fluctuations in both the plasma and magnetic field structure in the overlying helmet streamer belt. Additionally, regions of the streamer belt/HCS that are highly inclined with respect to the ecliptic are, in general, likely experiencing continual interchange reconnection to preserve the rigid rotation of coronal holes and their low-latitude extensions \citep{WangYM2004, Lionello2005}.
}

\subsection{{Intermittent Streamer Blob Outflow as a Source of Heliospheric Turbulence}}
\label{sec:disc:turb} 

{
Direct, in-situ magnetic field and plasma measurements by \textsl{Ulysses}, ISEE~3, and other spacecraft of small flux rope structures and the structured variability in density and other solar wind properties in the plasma sheet region surrounding the HCS led \citet{Crooker1996hps,Crooker1996hfds,Crooker2004} to interpret this region as being essentially filled with tangled, slow solar wind flux rope or flux tube transients. Heavy ion elemental and ionic composition measurements are consistent with this scenario, showing statistical properties of time series with lots of discontinuities \citep[e.g.,][]{Zurbuchen2002} and coincident boundaries in magnetic field, bulk plasma, and composition quantities \citep[e.g.,][and references therein]{Kepko2016,Viall2020}.
} 

{
As discussed by \citet{Higginson2018}, a major implication of the heliospheric plasma sheet region being essentially filled with tangled and possibly interacting small flux ropes, is the role they play in the development of solar wind turbulence \citep[e.g.,][]{Zheng2018,ZhaoLL2020}. Future work on this aspect of the slow solar wind's structured variability could include: (1) investigating the evolutionary processes that streamer blob/plasmoid flux ropes and other reconnection-generated outflows experience during their heliospheric transit \citep[e.g.,][]{Borovsky2012, Janvier2014a, MurphyA2020}; (2) studying how this variability interacts with stream interaction regions or coronal/heliospheric transients such as CMEs or CME-driven shocks and sheath regions \citep[e.g.,][]{Borovsky2006, Malandraki2019, Good2020};  (3) determining the contribution of magnetic island turbulence to particle energization processes in and around coronal streamers, the slow solar wind, and/or the HCS and plasma sheet \citep[e.g.,][]{Drake2006b, Dahlin2014, Guidoni2016, Khabarova2016}; and (4) whether this variabilty is sufficient to generate the seed population necessary for the most intense shock-accelerated SEP profiles \citep[e.g.,][]{Dahlin2015, Dahlin2016, Khabarova2015, Khabarova2016}. 
}

\section{Summary and Conclusions}
\label{sec:summary} 

This work presents an analysis of coronal inflows, generated as a consequence of intermittent reconnection at the cusp of helmet streamer belt and base of the HCS, in a time-dependent MHD simulation of a quasi-steady state solar wind in the moderately complex global magnetic field configuration of CR 2165 in July 2015. The simulation results confirm that the different viewpoints/orientations of the line-of-sight integration of Thomson-scattered white light give rise to different inflow morphologies. The dark, collimated `sinking column'/`tadpole-like' inflows are seen at sector boundaries when the HCS is face-on with respect to the observer. The semi-circular `shrinking loop' inflows are seen when the streamer belt arcade is oriented edge-on. Significantly, these two different types of inflows are indistinguishable from their height--time or velocity profiles alone, strongly suggesting the same underlying physical mechanisms---consistent with the prevailing interpretation of a common reconnection process.

While the \citet{Higginson2018} analysis was concentrated on the magnetic field and plasma signatures of the streamer blob flux ropes generated in an idealized equatorial HCS configuration, this paper focused on the low-corona signatures of the same magnetic reconnection processes. Taken together, these MHD simulations of dynamic, time-dependent coronal streamer evolution in a quasi-steady state slow solar wind represent a significant step forward in terms of begining to capture some of the intrinsic variability of the slow solar wind with numerical models. Understanding the origin and evolution the magnetic field and plasma signatures associated with streamer blob flux ropes, coronal downflows, in/out pairs, and streamer detachments will be an increasingly necessary component of the interpretation and analysis of current and future PSP and SolO observations. These data are expected make a significant contribution to the ability to observe and model the direct connection between heliospheric in-situ measurements and their origin in the solar corona.

\acknowledgments

B.J.L. acknowledges support from the NASA HGI and HSR programs 80NSSC18K0645, 80NSSC18K1553, and 80NSSC20K1448. This work utilizes GONG data from NSO, which is operated by AURA under a cooperative agreement with NSF and with additional financial support from NOAA, NASA, and USAF. The computational resources for this work were provided by the NASA High-End Computing Program through the NASA Center for Climate Simulation at Goddard Space Flight Center.

\bibliography{ms_arxiv}{}
\bibliographystyle{aasjournal}



\end{document}